\documentclass[12pt]{article}

\usepackage{amsfonts,amssymb,amsmath,latexsym,epsfig,amsthm,graphicx,euscript}

\newcommand{\pd}{\partial}

\def\kk{k}
\newcommand{\Dd}{\EuScript{D}^2}
\newcommand{\Dh}{\Dd_\mathrm{H}}
\newcommand{\DNh}{\tilde{\EuScript{D}}^2_\mathrm{H}}

\title{
\vskip -50pt
{\begin{normalsize}
\mbox{} \hfill DAMTP-2008-23 \\
\vskip 50pt
\end{normalsize}}
Rolling Solution for Tachyon Condensation\\ in Open String Field Theory
}

\author{Liudmila Joukovskaya\thanks{E-mail: \texttt{l.joukovskaya@damtp.cam.ac.uk}}\\
DAMTP, Centre for Mathematical Sciences, University of Cambridge,\\
Wilberforce Road, Cambridge CB3 0WA, UK}

\date{}

\begin{document}

\maketitle

\begin{abstract}
Open string field theory in the level truncation approximation is considered.
It is shown that the energy conservation law determines the existence of rolling
tachyon solution. The coupling of the open string field theory action to a
Friedmann-Robertson-Walker metric is considered and as a result the new time
dependent rolling tachyon solution is presented and
possible cosmological consequences are discussed.
\end{abstract}

\section{Introduction}
Consideration of fundamental theories such as M/String Theory
in the cosmological context continues to attract attention in the literature.

One of the interesting questions is the role of the tachyon in String
Theory and Cosmology. The great progress in our understanding about tachyon condensation
was made in the past decade\cite{Sen2002,Sen2004} but a lot of interesting issues
are still open. Among the most important ones is a better understanding of the dynamics
in tachyon condensation process.

Open String Field Theory (OSFT) \cite{Witten} gives us a tachyon
effective action\cite{Kost-Sam,MSZ} which is derived from first principles
and correctly describes tachyon physics and could represent most
exhaustive framework for studying tachyon dynamics.

The tachyon dynamics even in the case without gravity represents a nontrivial
task which was widely studied \cite{MZ,FGN,CMN,FH,HS}, while considering OSFT coupled to
the gravity in a Friedmann-Robertson-Walker (FRW) type of metric makes this problem
more nontrivial because of the d'Alambert operators in curved spaces that would appear
in the action. This fact makes  mathematical investigation
of this problem much more complicated but will benefit us with  desired time-dependent solution
interpolating between different vacua.

In this work new numerical rolling solution for tachyon condensation will be presented.
We will show that consideration of OSFT coupled to the gravity which represents more realistic
situation from cosmological point of view allows the existence of
the rolling tachyon configuration which was forbidden by energy conservation
law in the case without gravity.

\section{The Model}
The action of bosonic cubic string field theory has the form
\begin{equation}
S=-\frac{1}{g_0^2}\int(\frac{1}{2}\Phi \cdot Q_B \Phi+\frac{1}{3} \Phi \cdot (\Phi \ast \Phi)),
\label{action-full}
\end{equation}
where $g_0$ is the open string coupling constant, $Q_B$ is BRST operator,
$\ast$ is noncommutative product and $\Phi$ is the open string field containing
component fields which correspond to all the states in string Fock space.

Considering only tachyon field $\phi(x)$  at the level (0,0)
the action (\ref{action}) becomes
\begin{equation}
S=\frac{1}{g_0^2}\int d^{26}x\left[
\frac{\alpha^{\prime}}{2}\phi(x)\square\phi(x)+\frac{1}{2}\phi^2(x)
-\frac{1}{3}K^3 \Phi^3(x)-\Lambda
\right],
\label{action}
\end{equation}
where  $\alpha^{\prime}$ is the Regge slope,
$K=\frac{3 \sqrt{3}}{4}$, $\phi$ is a scalar field, $\Phi=e^{k\square_g}\phi$,
$k=\alpha^{\prime} \ln K$, $\square=\frac1{\sqrt{-g}}\pd_{\mu}\sqrt{-g}g^{\mu\nu}\pd_{\nu}$
and  $\Lambda =\frac{1}{6} K^{-6}$ was added to the potential to set
the local minimum of the potential to zero according Sen's conjecture \cite{Sen1999}.\\
The action (\ref{action}) leads to equation of motion
\begin{equation}
(\alpha^{\prime} \square+1)e^{-2 k \square} \Phi=K^3 \Phi^2.
\label{eom}
\end{equation}
The Stress Tensor for our system is\footnote{Note that here and below integration over $\rho$
is understood as limit of the following regularization
$$
\int_0^1d \rho f(\rho)= \lim _{\epsilon_1\to +0} \lim _{\epsilon_2\to +0}
\int^{1-\epsilon_2}_{\epsilon_1} d \rho f(\rho).
$$}
\begin{equation}
T_{\alpha\beta}(x)=
-g_{\alpha\beta}
\left(
    \frac{1}{2}\phi^2
    -\frac{\alpha^{\prime}}{2}\partial_\mu\phi\partial^\mu\phi
    -\frac{1}{3}K^3\Phi^3-\Lambda
\right)
-\alpha^{\prime} \partial_\alpha\phi\partial_\beta\phi
\end{equation}
$$
-g_{\alpha\beta}\,\kk
\int_0^1 d\rho
\left[
    (e^{k\rho\square}K^3 \Phi^2)
    (\square e^{-\kk\rho\square}\Phi)
    +
    (\partial_{\mu}e^{\kk\rho\square}K^3 \Phi^2)
    (\partial^{\mu}e^{-\kk\rho\square}\Phi)
\right]
$$
$$
+2\kk\int_0^1d\rho~
(\partial_{\alpha}e^{\kk\rho\square}K^3 \Phi^2)
(\partial_{\beta}e^{-\kk\rho\square}\Phi).
$$
The energy is defined as $E(t)=T^{00}$
and pressure as $P(t)_i=-T_i^i$ (no summation) and for our system are
$$
\mathcal{E}=\mathcal{E}_k+\mathcal{E}_p+\Lambda
+\mathcal{E}_{nl1}+\mathcal{E}_{nl2},~~~
\mathcal{P}=\mathcal{E}_k-\mathcal{E}_p-\Lambda
-\mathcal{E}_{nl1}+\mathcal{E}_{nl2}
$$
where
$$
{\cal E}_{k}=\frac{\alpha^{\prime}}{2}(\pd \phi)^2,~~~
{\cal E}_{p}=-\frac{1}{2}\phi^2+\frac{K^3}{3}\Phi^3
$$
$$
{\cal E}_{nl1}=k\int_{0}^{1}d \rho \left(
e^{k\rho \square} K^3 \Phi^2 \right)
 \left(-\square e^{-k\rho\square}\Phi\right),
$$
$$
{\cal E}_{nl2}=-k\int_{0}^{1} d \rho \left(\partial
e^{k\rho \square}K^3 \Phi^2\right) \left(\partial e^{-k \rho \square}\Phi\right).
$$
In this paper we will be interested in  spatially homogeneous configurations
for which Beltrami-Laplace operator used above takes the form $\square_g=-\pd^2$.
To avoid calculation of $e^{-k \rho\pd^2}$ term which
is much harder to compute then $e^{k \rho\pd^2}$ ($k>0$) as computation of the former
results in an ill-posed problem we will use the following representation for
nonlocal energy terms  ${\cal E}_{nl1}$ and ${\cal E}_{nl2}$  which are valid on the equation of motion
for the scalar field
$$
{\cal E}_{nl1}=k\int_{0}^{1}d \rho \left((-\alpha^{\prime} \pd^2+1)
e^{(2-\rho)k \pd^2}\Phi \right)
 \left(\pd^2 e^{k\rho\pd^2}\Phi\right),
$$
$$
{\cal E}_{nl2}=-k\int_{0}^{1} d \rho \left(\partial (-\alpha^{\prime} \pd^2+1)
e^{(2-\rho)k\pd^2}
\Phi\right) \left(\partial e^{k \rho \pd^2}\Phi\right).
$$
\textbf{Claim 1}\footnote{Similar theorem was proved in \cite{MZ}, but the
variant we use is more useful for numerical calculations, because in order
to define action of the exponential operator we need to do only one well defined integration
instead of the summation over infinite series expansions with which one always need
to be very careful about the convergence and related issues.}
The Energy
$$
E=\frac{\alpha^{\prime}}{2}(\pd\phi)^2-\frac{1}{2}\phi^2+\frac{K^3}{3}\Phi^3+\Lambda+
\kk\int^1_0 d \rho~((-\alpha^{\prime} \pd^2 +1) e^{(2-\rho) \kk \pd^2} \Phi)
\overleftrightarrow{\partial}(\partial e^{\kk\rho\pd^2} \Phi),
$$
is conserved on the solutions of equation of motion (\ref{eom})
$$
(-\alpha^{\prime} \partial^2+1)e^{2\kk\pd^2}\Phi=K^3 \Phi^2
$$
where $A\overleftrightarrow{\partial}B=A\pd B-B \pd A$.\\
\textbf{Proof.}
$$
\frac{d E(t)}{dt}=\alpha^{\prime} \pd\phi\pd^2 \phi-\phi\pd
\phi+\Phi^3\pd \Phi+
k\int^1_0 d \rho((-\alpha^{\prime} \pd^2 +1)e^{(2-\rho)k\pd^2}\Phi)
\overleftrightarrow{\partial^2}
(\partial e^{k\rho\pd^2}\Phi).
$$
Using following identity\cite{Lulya-paper-2-ext}
$$
\int\limits_0^1 d\rho (e^{ \rho
\partial^2 }\varphi ) \overleftrightarrow{\partial^2} (e^{ (1
- \rho)\partial^2 } \phi)= \varphi \overleftrightarrow {e^{\partial^2 }}\phi,
$$
equation of motion and definition of field $\Phi$, we have
$$
\frac{d E(t)}{dt}=\alpha^{\prime} \pd \phi \pd^2 \phi-\phi\pd
\phi+\Phi^3
 \pd \Phi+\partial  \Phi\overleftrightarrow{e^{k\pd^2}}
 (\alpha^{\prime}\pd^2-1)e^{k\pd^2}\Phi=\pd \Phi\left[\Phi^3+(\alpha^{\prime} \partial^2-1)e^{2k \pd^2}\Phi
\right]=0.
$$
It is straightforward to generalize this statement to arbitrary potential and
only finite number of fields.
Let us consider physical consequence of the energy conservation law.\\
\textbf{Claim 2}\footnote{The similar claim for the p-adic string model
was proved in \cite{MZ}, which  rules out the possibility that the tachyon may roll monotonically down
from one extremum reaching the tachyon vacuum.}
There doesn't exist continuous solution of equation (\ref{eom}) which satisfies
the boundary conditions
\begin{equation}
\label{bc}
\lim \Phi(t)=
\begin{cases}
0,&t\to\infty,\\
K^{-3},&t\to-\infty
\end{cases}
\end{equation}
or vice-versa (in terms $t \to -t$).\\
\textbf{Proof}.
Let us assume existence of such solution and calculate energy
at the extremum  points, we get  $E(\Phi=0)=\Lambda$
and $E(\Phi=K^{-3})=-\frac{1}{6}K^{-6}+\Lambda$, i.e. energy values at
$t \to +\infty$  and  $t \to -\infty$ are different what contradicts the energy conservation
theorem.

As we can see energy conservation law plays crucial role in the existence of the time dependent
solutions of equation of motion for the case of level truncation approximation for OSFT.
The above statement could probably be generalized to the case of full OSFT action because for
the action with cubic interaction solution interpolating between maximum and minimum
 in the effective potential has to interpolate between vacua with different energy.

\section{The Model Coupled to the Gravity}
\label{rolling-tachyon}
In this section we  consider more realistic case when gravity is included
\begin{equation}
\label{action-gr}
S=\frac{1}{g_0^2} \int d^4x\sqrt{-g}\left(\frac{m_p^2}{2}R+
\frac{1}{2}\phi \square_g \phi
+\frac{1}{2}\phi^2-\frac{1}{3}K^3\Phi^3-\Lambda \right),
\end{equation}
here $m_p^2=g_0^2 M_{pl}^2$ and we will work in units where $\alpha^{\prime}=1$.
As a particular metric we will consider a FRW one
$$
ds^2=-dt^2+a^2(t)(d x_1^2+dx_2^2+dx_3^2),
$$
for which the Beltrami-Laplace operator for spatially-homogeneous configurations
takes the form $\square_g=-\pd^2-3 H(t) \pd=-\Dh$.
Scalar field and Friedmann equations are
\begin{equation}
(-\Dh+1)e^{2k \Dh}\Phi = K^3 \Phi^2,
\label{eom-gr}
\end{equation}
\begin{equation}
3H^2=\frac{1}{m_p^2}~{\cal E},~~~
3H^2+2\dot H={}-\frac{1}{m_p^2}~{\cal P}.
\label{Fr-eq}
\end{equation}
Inclusion the gravity drastically changes the question of existence
of the dynamical interpolation between maximum and minimum of the scalar field potential.
This happens because then there are no restrictions from energy conservation law.

According to the Friedmann and scalar field equations we can expect
the scalar field rolling solution and Hubble function satisfying the
following boundary conditions
\begin{equation}
\label{bc-gr}
\lim \Phi(t)=
\begin{cases}
0,&t\to\infty,\\
K^{-3},&t\to-\infty
\end{cases}
~~~~~~~
\lim H(t)=
\begin{cases}
(18 K^6)^{-1/2},&t\to\infty,\\
0,&t\to-\infty
\end{cases}
\end{equation}
or vice-versa (in terms of $t \to -t$). Note that from cosmological perspective
we are interested only in positive values for the Hubble function
and thus we do not consider negative sign in front of the square root in (\ref{bc-gr}).

\begin{figure}
\centering
\includegraphics[width=71mm]{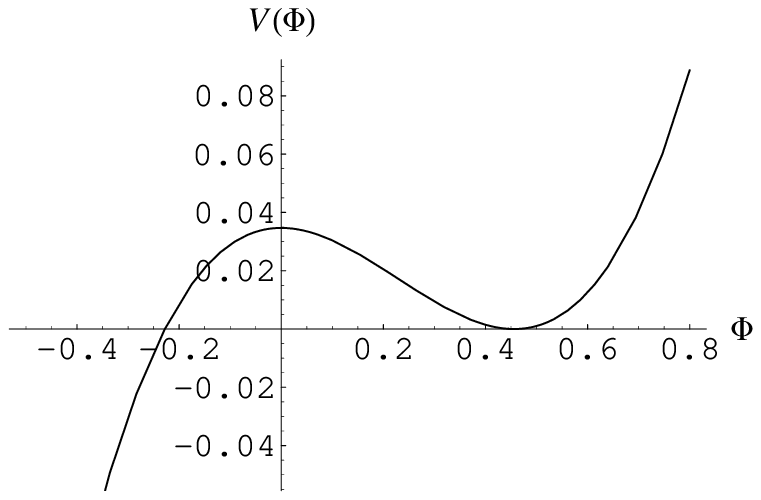}
\caption{The potential}
\label{cubic-potential}
\end{figure}

To analyze physical situation let us consider potential  in which motion is expected.
Naive extraction of potential from the model action (\ref{action-gr}) results in
$V(\Phi)=-\frac{1}{2}\Phi^2+\frac{1}{3}K^3 \Phi^3+\Lambda$. The constant  $\Lambda$
represents the D-brane tension which according to Sen's conjecture must be added to cancel the
negative energy appearing due to the presence of tachyon. We have obtained two type of solutions.
The first one is an ordinary rolling solution which starts from $\Phi=0$ and goes
towards configuration $\Phi=K^{-3}$ which is associated with the true vacuum.
This solution can be interpreted as a description of the D-brane decay.
The second one is a rolling solution which goes in the opposite direction,
which appears in this model possibly because of the non-locality in the interaction.
It is known that nonlocal dynamics has many interesting properties which are not present
in the local case. In particular the ``slop effect'' \cite{MZ,FGN,AJ} which is present
in the obtained solutions (Fig. \ref{rolling-solution-ord}, \ref{rolling-solution-ph})
when the scalar field goes beyond the values from which the scalar field configuration starts --
situation which is not possible in the local models.
Potentially a similar effect can initiate non-symmetry in the potential in
ekpyrotic \cite{Ekpyrotic} and cyclic cosmology \cite{Cyclic}.

\section{Numerical Scheme for Solution Construction}
For numerical calculations we operate with scalar field equation of motion (\ref{eom-gr})
and the difference of equations (\ref{Fr-eq})
\begin{equation}
\label{fr2}
(-\Dh+1)e^{2k \Dh}\Phi = K^3 \Phi^2,~~
\dot{H}=-\frac{1}{2m_p^2}~(\cal P+\cal E).
\end{equation}
The outline of the numerical scheme is the following\footnote{
It is important that the described below scheme is general and was also used for
solving cosmological equations for the case of Cubic Fermionic Field Theory with
the quartic interaction term in (\ref{action-gr}) instead of cubic one as well as for
p-adic string model at least for $p=2,3$. Obtained interpolating solutions
between corresponding maximum and minimum of the tachyon potential look very similar
to one which will be presented in section \ref{Rolling-Tachyon-solution} and will be presented in
\cite{Lulya-paper-2-ext}. It looks that cosmology effaces difference between
cubic and quartic interaction for the type of solutions indicated above. }

\begin{itemize}
\item
For equations (\ref{fr2}) we introduce lattice in $t$ variable and
then solve resulting system of nonlinear equations using iterative
relaxation solver using discrete $L_2$ norm to control error
tolerance.

\item
The nontrivial thing from computational point of view is efficient evaluation of
$e^{2k\rho\Dh}\Phi$ for $\rho\in[0,2]$.
This operator could be interpreted in terms of initial value problem
for the following diffusion equation with boundary conditions
\begin{equation}
\label{numdiff}
\partial_\rho\varphi(t,\rho)=
\partial^2_t\varphi(t,\rho)+3H(t)\partial^2_t\varphi(t,\rho),
\end{equation}
$$
\varphi(0,t)=\Phi(t),~~\varphi(\rho,\pm\infty)=\Phi(\pm\infty).
$$
Once solution of this equation is constructed we have $e^{2k\rho\Dh}\Phi(t)=\varphi(\rho,t)$.

\item
To solve (\ref{numdiff}) we used second order Crank-Nicholson scheme which is based on
approximation
$$
e^{2k\Delta_\rho\DNh}\varphi=
\left(1+k\Delta_\rho\DNh\right)\left(1-k\Delta_\rho\DNh\right)^{-1}\varphi+
o(\Delta_\rho^2\|\DNh\|),
$$
where $\DNh$ is a $\Dh$ operator on the $t$-lattice (it thus has a
finite norm) and $\Delta_\rho$ is a step size along $\rho$ variable.
Derivatives in $t$ variable were approximated using 4th order finite
differences on uniform lattice (symmetric scheme).

\item
In order to exclude possible artifacts of this specific numerical
scheme we tried Chebyshev-pseudospectral method which is known to
have impressive exponential convergence \cite{Forn}. This scheme is
known to have very different properties \cite{Forn} compared to finite
difference scheme described above, but it produced the same results
up to the approximation error which gives us confidence in the
existence of the rolling solution reported in this work.
\end{itemize}

\section{Rolling Tachyon Solution and their Cosmological Consequences}
\label{Rolling-Tachyon-solution}

\begin{figure}
\centering
\includegraphics[width=51mm]{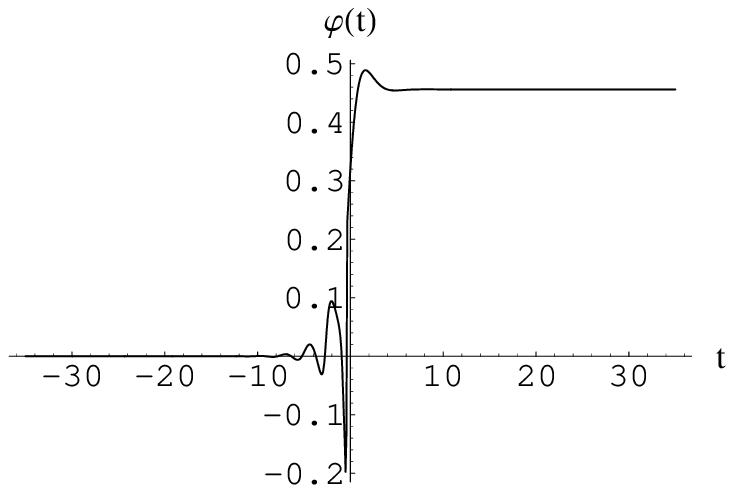}~
\includegraphics[width=51mm]{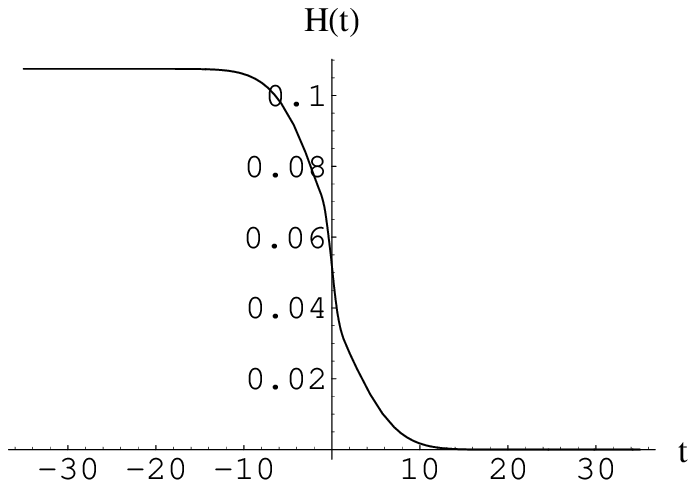}~
\includegraphics[width=51mm]{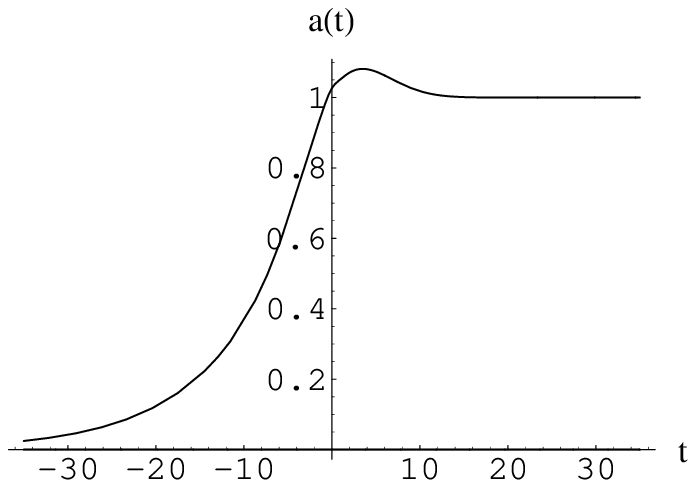}~
\caption{Solutions of the scalar field (\ref{eom-gr}) and  Friedmann
equation (\ref{Fr-eq}) $\Phi$, $H$ and $a$ (left to right) for
$m_p^2=1$.} \label{rolling-solution-ord}
\end{figure}

\begin{figure}[b]
\centering
\includegraphics[width=51mm]{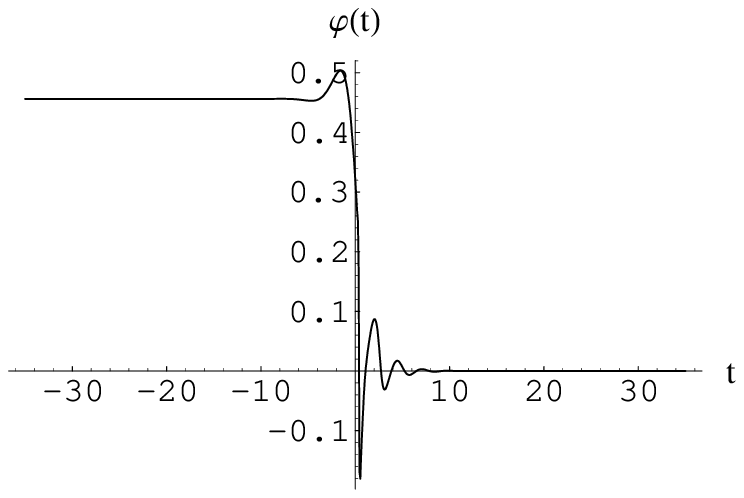}~
\includegraphics[width=51mm]{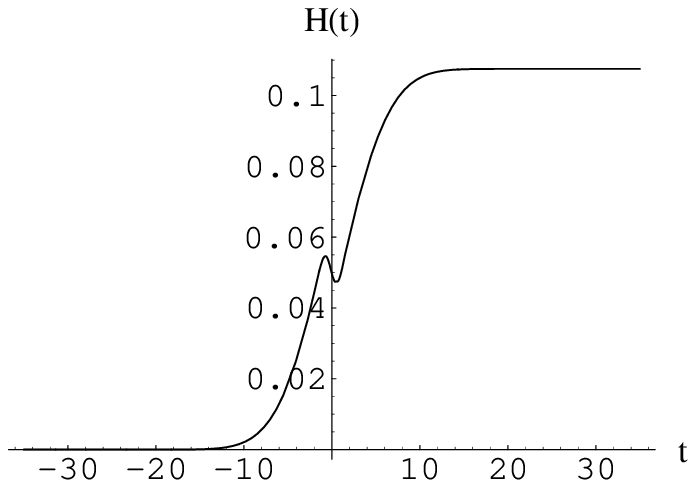}~
\includegraphics[width=51mm]{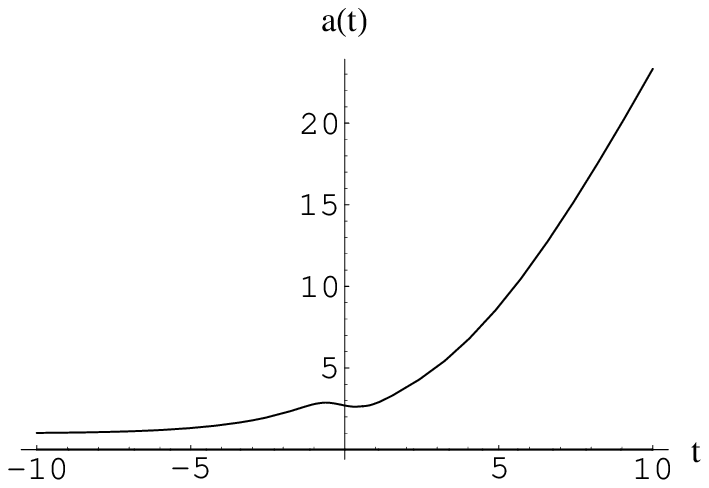}~
\caption{Solutions of the scalar field (\ref{eom-gr}) and  Friedmann
equation (\ref{Fr-eq}) $\Phi$, $H$ and $a$ (left to right) for
$m_p^2=1$.} \label{rolling-solution-ph}
\end{figure}

Solutions of (\ref{eom-gr}) and (\ref{Fr-eq}) are presented
on Fig. \ref{rolling-solution-ord} and \ref{rolling-solution-ph}.
As we can see we obtained accelerating rolling solutions for tachyon scalar field $\Phi$.
It is natural to address cosmological issues in the context of
rolling tachyon solution \cite{Gibbons}. Taking into account
that acceleration of the Universe is one of the most fascinating
processes of the modern cosmology many authors tried to explain
among other possibilities with a possible explanation via scalar
field. It is interesting that cosmology gives us this solution owing
to coupling our action (\ref{action}) to a FRW metric and the
consequent inclusion of a Hubble friction term which leads to
time-dependent rolling solution with exponentially decreasing
oscillations around the minimum. Moreover because generally speaking
string scale does not exactly coincide with Plank mass we obtain
some freedom in settling $m_p^2$ parameter for numerical
calculations which enters into Friedmann equations and as a result
govern the value of Hubble function $H(t)$. Thus
decreasing the value of $m_p^2$ leads to more smooth profile for rolling
solution while increasing $m_p^2$ results in higher
oscillations of the solution in comparison to those presented on the
Fig. \ref{rolling-solution-ord} and \ref{rolling-solution-ph}, more details will
be presented in
\cite{Lulya-paper-2-ext}. During the process of completion of this work appeared \cite{HS}
in which OSFT tachyon in the dilaton background was considered and
time-like rolling tachyon solution  were obtained. Because dilaton
 appears from the same string sector as graviton
including the dilaton into the tachyon action can qualitatively
reproduce behavior of the tachyon in the curved spaces.

Concluding this section we would like to summarize that we obtained time dependent
accelerating solution interpolating between unstable
and the true vacua (see Fig. \ref{rolling-solution-ord}) which can be interpreted as
being responsible for acceleration of the Universe during this rolling from unstable
vacuum to the true vacuum, after which it disappears.
Evolution of the scalar field in the opposite direction is also possible with
the Hubble function in form of increasing kink when scale factor $a(t)$ starts
from the constant plateau and exponentially grows, which seems
counter intuitive but can be related to late time acceleration.

\section{Conclusion}
The Witten's cubic open bosonic string filed theory in the level truncation
approximation was considered. It was shown that the energy conservation law determines
existence of rolling tachyon solution. As a result it was explicitly shown that
the non-existence of the rolling solution in the Minkowski case is a necessary consequence
of the energy conservation law of the system. The modification of conservation
law in the presence of the gravity  is discussed.
The first rolling solution for tachyon condensation in this theory is presented and
possible cosmological consequences are discussed.  Although
only lowest excitation in the full OSFT were taken into account there are solid
reasons to suppose that the general picture for the tachyon condensation process
will be the same in the case of full OSFT.

\section*{Acknowledgements}
The author would like to thank I. Aref'eva, R. Bradenberger, A.-C. Davis,
J. Khoury, N. Nunes,  F. Quevedo, D. Seery, D. Wesley and especially
D. Mulryne and  Ya. Volovich for  useful discussions.
The author gratefully acknowledge the use of the UK National
Supercomputer, COSMOS, funded by PPARC, HEFCE and Silicon
Graphics. This work is supported by  the Centre for Theoretical
Cosmology, in Cambridge.

\end{document}